\PassOptionsToPackage{table,xcdraw}{xcolor}
\documentclass[11pt,a4paper]{article}
\pdfoutput=1
\usepackage{jheppub}
\usepackage{gensymb}
\DeclareSymbolFont{matha}{OML}{txmi}{m}{it}
\DeclareMathSymbol{\varv}{\mathord}{matha}{118}
\usepackage{amsmath}
\usepackage{amssymb}
\usepackage{graphicx}
\usepackage{subfigure}
\usepackage{pstricks}
\usepackage{bm}
\usepackage{pbox}
\usepackage{placeins}
\usepackage{booktabs}
\usepackage[T1]{fontenc}
\usepackage{footnote}
\usepackage{pdfpages}
\usepackage{hhline}
\usepackage{multirow}
\usepackage{multicol}
\usepackage{enumitem}
\usepackage[toc,page]{appendix}
\allowdisplaybreaks
\usepackage{comment}
\usepackage{float}
\usepackage{slashed}
\usepackage{xcolor}
\usepackage{textcomp}
\usepackage{gensymb}
\usepackage{multirow}
\usepackage[numbers]{natbib}
\usepackage{notoccite}
\usepackage{cancel}
\usepackage{rotating}
\usepackage{tabularx}
\usepackage[labelfont=bf]{caption} 
\usepackage{soul}

\renewcommand{\thetable}{\Roman{table}}

\renewcommand{\tableofcontents}{}

\definecolor{MyDarkBlue}{rgb}{0.1, 0.1, 0.8} 
\definecolor{MyLightBlue}{rgb}{0.22,0.51,0.9}
\definecolor{MyGreen}{rgb}{0.0, 0.5, 0.0}
\definecolor{BrickRed}{rgb}{0.8, 0.25, 0.33}

\RequirePackage{hyperref}
\definecolor{MyLinkColor}{RGB}{180,60,90}
\hypersetup{
  colorlinks,
  citecolor=blue,
  linkcolor=MyLinkColor,
  urlcolor=MyLightBlue
}
\makeatletter
\gdef\@fpheader{}
\makeatother
\begin{document}
\title{\bf 
Towards Precision Neutrino Fits in GUTs: Relevance of One-Loop Finite Corrections
}

\author[a]{Chee Sheng Fong,}
\author[b]{Shaikh Saad}  

\vspace{0.5cm}

\affiliation[a]{Centro de Ciências Naturais e Humanas,
Universidade Federal do ABC, 09.210-170,\\ Santo André, SP, Brazil}

\affiliation[b]{Jožef Stefan Institute, Jamova 39, P.\ O.\ Box 3000, SI-1001 Ljubljana, Slovenia}

\emailAdd{sheng.fong@ufabc.edu.br, shaikh.saad@ijs.si}
\abstract{
In this work, we perform a dedicated analysis of fermion mass fits in the minimal $SO(10)$ grand unified theory (GUT), going beyond the tree-level approximation by incorporating one-loop finite corrections to the neutrino mass matrix. We show that parameter regions that successfully reproduce all fermion masses and mixings at tree level can lead to significant deviations in neutrino masses and leptonic mixing parameters once the radiative corrections are included. These results expose a limitation of conventional tree-level fitting procedures and highlight the sensitivity of neutrino observables to loop effects. Since in the minimal $SO(10)$ GUT the same set of Yukawa parameters simultaneously governs quark masses, charged lepton masses, and neutrino properties, these radiative corrections propagate across all fermion sectors, reshaping the viable parameter space in a highly non-trivial and correlated manner. We find that the largest corrections to the masses and mixing angles are of order $\mathcal{O}(30\%)$-$\mathcal{O}(40\%)$, therefore, cannot be neglected. In light of the current precision of neutrino oscillation measurements, and the expected improvements from ongoing and future experiments, we demonstrate that the inclusion of one-loop effects is essential for a consistent and reliable exploration of the parameter space, with important implications for the predictivity of $SO(10)$ GUTs.
}

\maketitle

\section{Introduction}
Grand unified theories (GUTs)~\cite{Pati:1974yy, Georgi:1974sy, Georgi:1974yf, Georgi:1974my, Fritzsch:1974nn} based on the $SO(10)$ gauge group provide a particularly well-motivated framework for physics beyond the Standard Model, as they unify all fermions of a single generation into a single irreducible representation that naturally includes a right-handed neutrino. This feature offers a compelling explanation for the origin of neutrino masses via the Type-I and Type-II seesaw mechanisms~\cite{Minkowski:1977sc,Yanagida:1979as,Glashow:1979nm,Gell-Mann:1979vob,Mohapatra:1979ia,Schechter:1980gr,Schechter:1981cv}. Moreover, quark–lepton unification implies that all fermion masses and mixings arise from a common set of Yukawa couplings, leading to highly constrained and predictive structures. As a result, fermion mass fits in $SO(10)$ models provide a powerful probe of the underlying theory, linking observables across the quark and lepton sectors. A significant effort has been devoted to obtaining a realistic description of the fermion mass spectrum within renormalizable $SO(10)$ GUTs.
Both non-supersymmetric and supersymmetric realizations have been explored in Refs.~\cite{Babu:1992ia, Joshipura:2011nn, Dueck:2013gca,Babu:2016cri, Babu:2020tnf}.
Analyses concentrating specifically on the non-supersymmetric case can be found in Refs.~\cite{Altarelli:2013aqa, Babu:2016bmy,Ohlsson:2018qpt, Ohlsson:2019sja,Mummidi:2021anm,Fu:2022lrn, Saad:2022mzu,Haba:2023dvo, Kaladharan:2023zbr,Babu:2024ahk,Fong:2025aya, Babu:2025wop,Chen:2025afg, Saad:2026ucf},
while purely supersymmetric constructions have been studied in Refs.~\cite{Babu:1995fp, Bajc:2001fe,Bajc:2002iw, Fukuyama:2002ch,Goh:2003sy, Goh:2003hf,Dutta:2004hp,Bertolini:2004eq, Bertolini:2005qb,Babu:2005ia,Dutta:2005ni, Bertolini:2006pe,Aulakh:2006hs,Grimus:2006rk,Bajc:2008dc,Fukuyama:2015kra, Babu:2018tfi,Babu:2018qca, Mohapatra:2018biy,Mimura:2019yfi,Fu:2023mdu}.

All existing analyses of fermion masses in this framework are performed at the tree level, where successful fits to charged as well as neutral fermion masses and mixing observables can be readily obtained. However, the neutrino sector is particularly sensitive to subleading effects due to the seesaw structure, raising the question of the reliability of such tree-level approximation. In this work, we perform a detailed analysis of fermion mass fits in $SO(10)$ models beyond the tree-level approximation by incorporating finite one-loop corrections to the neutrino mass matrix, which are only smaller by a factor of order $(16\pi^2)^{-1}\ln(\Lambda_{\rm seesaw}/v)$ than the tree-level contribution 
and hence they are especially relevant for high seesaw scale $\Lambda_{\rm seesaw}$ in comparison to the weak scale $v$~\cite{Grimus:1989pu,Grimus:2002nk,AristizabalSierra:2011mn}.
We show that parameter regions consistent with charged fermion masses and quark mixing at tree level can lead to significant deviations in neutrino masses and leptonic mixing parameters once radiative corrections are included. In particular, loop effects can induce non-negligible shifts that spoil agreement with current neutrino oscillation data, thereby exposing a limitation of conventional fitting procedures. This effect is particularly significant in the $SO(10)$ framework, where a unified Yukawa structure underlies all fermion masses and mixings. Consequently, modifications arising in the neutrino sector are inherently linked to the rest of the fermion spectrum, leading to correlated shifts in the parameter space. 

Given that the measurements of neutrino oscillation parameters have now entered a precision era, with steadily improving experimental sensitivities, it is essential to achieve a comparable level of accuracy on the theoretical side. Our results demonstrate that the inclusion of one-loop finite corrections is crucial for a consistent and reliable exploration of the parameter space, 
with important implications for the viability and predictivity of $SO(10)$ unified frameworks.
Although our analysis is performed within a specific model, the resulting conclusions are generic and applicable to a broader class of GUT frameworks that implement the type-I seesaw mechanism.

The paper is organized as follows. In Sec.~\ref{sec02}, we discuss the Yukawa sector of the minimal $SO(10)$ model and the implementation of radiative corrections in the neutrino sector. In Sec.~\ref{sec03}, we present a detailed numerical analysis demonstrating the importance of these corrections. Finally, we conclude in Sec.~\ref{sec04}.
In Appendix~\ref{app:fit_parameters}, we present the benchmark fits of the model without and with one-loop finite corrections to the neutrino mass matrix while in Appendix~\ref{app:lepton_couplings}, we present the parameters in lepton sector at various scales when heavy right-handed neutrinos are integrated out.

\section{Model}\label{sec02}
We consider the minimal Yukawa sector~\cite{Babu:2016bmy} of $SO(10)$ GUT, where the fermion masses arise from their couplings to Higgs fields transforming in the $\mathbf{10}_H$, $\mathbf{120}_H$, and $\overline{\mathbf{126}}_H$ representations. In this setup, the $\mathbf{10}_H$ and $\mathbf{120}_H$ representations are taken to be real, while the $\overline{\mathbf{126}}_H$ is intrinsically complex. This minimal field content is sufficient to generate realistic fermion masses and mixings, while maintaining a predictive~\cite{Saad:2026ucf} framework due to the limited number of Yukawa structures.

After symmetry breaking, the fermion mass matrices can be expressed in terms of three independent Yukawa structures, denoted by $\mathcal{M}_i$ $(i=1,2,3)$, and a set of complex coefficients encoding the relative vacuum expectation values and phases. The resulting mass matrices take the form~\cite{Saad:2026ucf}
\begin{align}
&M_{U}= \mathcal{M}_1 + \mathcal{M}_2^\mathrm{diag} + \mathcal{M}_3, \label{eq:MU} \\
&M_{D}= \mathcal{M}_1+\hat p  \mathcal{M}_2^\mathrm{diag}+ q \mathcal{M}_3 , \\
&M_{E}= \mathcal{M}_1-3 \hat p  \mathcal{M}_2^\mathrm{diag}+ r \mathcal{M}_3 ,  \\
&M_{\nu_D}= \mathcal{M}_1-3  \mathcal{M}_2^\mathrm{diag}+ \frac{3q^*-r^*}{q^*+r^*} \mathcal{M}_3 \label{eq:MDnu}, \\
&M_{\nu_{R}} = s \mathcal{M}_2^\mathrm{diag}, \label{eq:MR}  
\end{align}
where $M_U$, $M_D$, $M_E$, and $M_{\nu_D}$ denote the up-quark, down-quark, charged lepton, and Dirac neutrino mass matrices, respectively, while $M_{\nu_R} = {\rm diag}(M_1,M_2,M_3)$ corresponds to the right-handed Majorana neutrino mass matrix. In general, the matrix $\mathcal{M}_1$ is complex symmetric and $\mathcal{M}_3$ is complex antisymmetric, whereas, in the chosen basis, $\mathcal{M}_2^\mathrm{diag}$ is real, positive, and diagonal.   
The coefficients determined by the ratios of  Higgs vacuum expectation values are three complex $\hat p$, $q$, $r$ and a real $s$.

The structure of these mass matrices reflects the underlying quark--lepton unification, leading to non-trivial correlations among fermion masses and mixing parameters. In particular, the presence of the $\overline{\mathbf{126}}_H$ representation generates the Majorana mass term Eq.~\eqref{eq:MR} for right-handed neutrinos, enabling the realization of the Type-I seesaw mechanism, 
leading to the following light neutrino mass matrix
\begin{align}
    M_N &= - M^T_{\nu_D} M^{-1}_{\nu_R} M_{\nu_D}. \label{eq:mnu}
\end{align}
We have chosen to neglect the type-II seesaw contribution since a type-II dominated scenario does not give a viable fit~\cite{Babu:2016bmy}.
We emphasize that Eqs.~\eqref{eq:MU}--\eqref{eq:MDnu}, together with Eq.~\eqref{eq:mnu}, are expressed in the right--left (RL) convention. Although the model allows for both normal and inverted neutrino mass ordering~\cite{Saad:2022mzu,Babu:2024ahk}, we focus on the normal mass ordering of light neutrinos for illustrative purposes.

\subsection{One-loop finite corrections to the neutrino masses}\label{D}
In the right-handed neutrino $N_i$ mass diagonal basis, the one-loop corrected neutrino mass matrix takes the following form~\cite{Grimus:2002nk,AristizabalSierra:2011mn},
\begin{align}
&M_N^\mathrm{1-loop} = M_N  + \delta M_L,
\label{eq:corrected_mnu}
\end{align}
where we have defined
\begin{align}
	&\delta M_L = M^T_{\nu_D} M^{-1}_{\nu_R} X M_{\nu_D}, \label{eq:1-loop}
	\\
	&X_{ij}=\frac{g_2^2\delta_{ij}}{64\pi^2M^2_W} \bigg\{
	\frac{M^2_h M_i^2}{M_i^2 - M_h^2} \ln \left( \frac{M^2_{i}}{M^2_h} \right)
	+3 \frac{M^2_Z M_i^2}{M_i^2 - M_Z^2} \ln \left( \frac{M^2_{i}}{M^2_Z} \right)
	\bigg\},\label{eq:X_function}
\end{align}
and take the following values:
$M_Z=91.1876\, \mathrm{GeV}$~\cite{Antusch:2025fpm}, 
$M_W=80.3692\, \mathrm{GeV}$~\cite{ParticleDataGroup:2024cfk},  $M_h=125.20\,\mathrm{GeV}$~\cite{ParticleDataGroup:2024cfk},  and $g_2=0.65096$~\cite{Antusch:2025fpm}.    

\begin{figure}\centering
	\includegraphics[width=0.65\columnwidth]{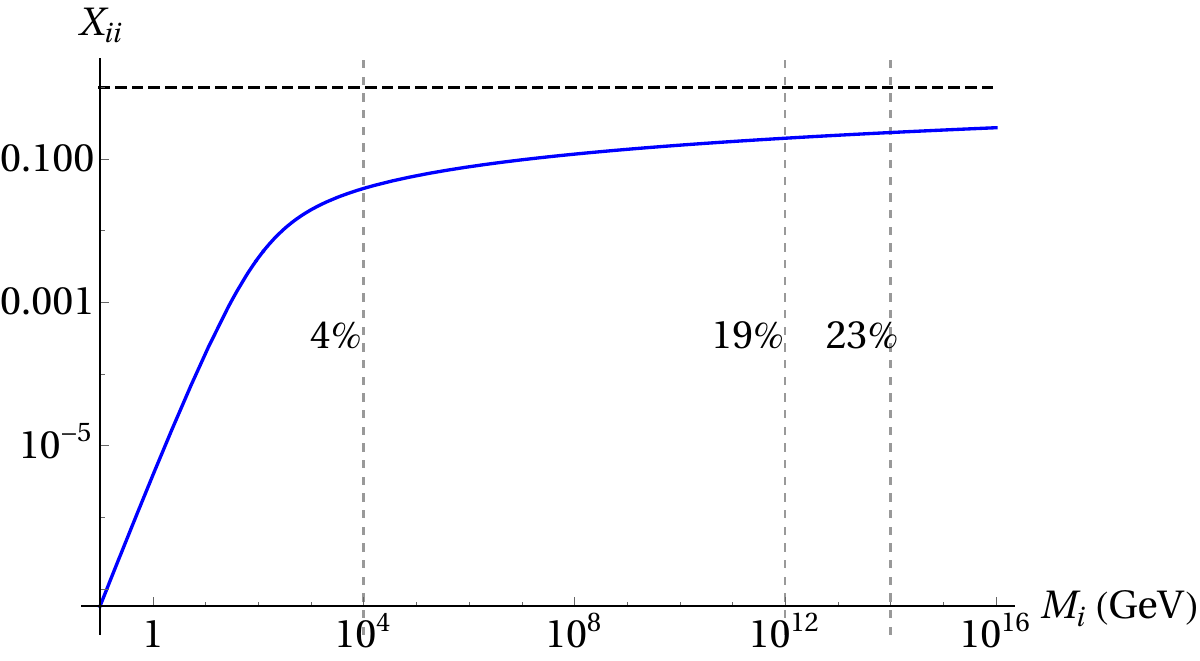}
	\caption{$X_{ii}$ defined in eq.~\eqref{eq:X_function} as a function of $M_i$. The black horizontal dashed line indicates unity while the vertical dashed lines show three representative values: $M_i = \{ 10^{4}, 10^{12}, 10^{14} \}$ GeV with their respective $X_{ii}$ values in percentage.}\label{fig:1loop_X}
\end{figure}

In Figure~\ref{fig:1loop_X}, we plot $X_{ii}$ as a function of $M_i$ to show the relative importance of the one-loop finite corrections.
In particular, the corrections become more important as $M_i$ increases while for $M_i < 100$ GeV, the corrections go below percent level,
although a recent analysis for low-scale seesaw in Ref.~\cite{Miele:2026uar} shows a deviation from this expectation. 
Notice that perturbation theory remains valid since $X_{ii}$ remains below one (the black horizontal dashed line) even when $M_i = 10^{16}$ GeV. 
We also show as the vertical dashed lines three representative values $M_i = \{ 10^{4}, 10^{12}, 10^{14} \}$ GeV with their respective $X_{ii}$ values in percentage $\{4\%,19\%,23\%\}$. 

It is important to note that $M_{\nu_D}$, to be used in Eq.~\eqref{eq:1-loop}, is not directly given by eq.~\eqref{eq:MDnu} due to the running through renormalization group equations (RGEs). Upon integrating out $N_i$, the corresponding elements $(M_{\nu_D})_{i\alpha}$ will be frozen. In order words, the $i$-row elements of $M_{\nu_D}$ are evaluated at the scale $\mu = M_i$ i.e. $(M_{\nu_D})_{i\alpha}(\mu = M_i)$. To be more specific, while computing $\delta M_L$ at the $M_Z$ scale, we use the frozen mass matrices for both $M_{\nu_D}$ and $M_{\nu_R}$. 
For further details, see Appendix \ref{app:lepton_couplings}.

Strictly speaking, for the tree-level neutrino mass matrix, the expression in eq.~\eqref{eq:mnu} is used only for energy scale above $M_3$ while at lower scale, the right-handed neutrinos are subsequently integrated out at their mass thresholds following the procedure of Ref.~\cite{Antusch:2005gp}.

\section{Numerical Analysis}\label{sec03}

We fit fermion masses and mixings starting from the GUT-scale relations in Eqs.~\eqref{eq:MU}--\eqref{eq:MR} and Eq.~\eqref{eq:mnu}, involving the 28 free parameters~\cite{Saad:2026ucf}. The Yukawa couplings and right-handed neutrino masses are evolved~\cite{Antusch:2005gp} from $M_{\rm GUT}=10^{16}\,\mathrm{GeV}$ down to $M_Z$ using SM+Type-I seesaw RGEs, with heavy neutrinos integrated out at their thresholds.  At $M_Z$, we
evaluate a $\chi^2$ function including charged fermion masses, CKM observables, and neutrino mass-squared differences and mixing angles. Experimental inputs in the charged fermion sector are taken from Ref.~\cite{Antusch:2025fpm}, while the neutrinos oscillation data are from  Refs.~\cite{NUFIT,Esteban:2020cvm}. For observables with experimental uncertainties below $1\%$, a minimum relative uncertainty of $1\%$ is imposed for numerical stability.

\begin{table}[t!]
\centering
\footnotesize
\resizebox{0.8\textwidth}{!}{
\begin{tabular}{lccccc}
\toprule

\textbf{Observable} & \textbf{Exp.} & \textbf{BM I} & \textbf{Pull} & \textbf{BM II} & \textbf{Pull} \\
\midrule

\multicolumn{6}{c}{\textbf{Up-type quarks}} \\
\midrule
$y_u \,(10^{-6})$ & $7.09^{+1.56}_{-0.88}$ & 6.889 & -0.23 & 6.940 & -0.17 \\
$y_c \,(10^{-3})$ & $3.55^{+0.10}_{-0.09}$ & 3.555 & +0.05 & 3.533 & -0.19 \\
$y_t$ & $0.968\pm0.004$ & 0.967 & -0.13 & 0.967 & -0.07 \\
\midrule

\multicolumn{6}{c}{\textbf{Down-type quarks}} \\
\midrule
$y_d \,(10^{-5})$ & $1.55^{+0.14}_{-0.07}$ & 1.550 & +0.00 & 1.549 & -0.02 \\
$y_s \,(10^{-4})$ & $3.10^{+0.26}_{-0.14}$ & 3.133 & +0.13 & 3.188 & +0.34 \\
$y_b \,(10^{-2})$ & $1.63^{+0.02}_{-0.01}$ & 1.635 & +0.27 & 1.632 & +0.09 \\
\midrule

\multicolumn{6}{c}{\textbf{Charged leptons}} \\
\midrule
$y_e \,(10^{-6})$ & $2.77705^{+0.00033}_{-0.00039}$ & 2.774 & -0.12 & 2.779 & +0.07 \\
$y_\mu \,(10^{-4})$ & $5.85026^{+0.00076}_{-0.00075}$ & 5.858 & +0.13 & 5.848 & -0.04 \\
$y_\tau \,(10^{-2})$ & $0.99370^{+0.00015}_{-0.00014}$ & 0.992 & -0.19 & 0.994 & +0.05 \\
\midrule

\multicolumn{6}{c}{\textbf{CKM parameters}} \\
\midrule
$\theta_{12}^q$ & $0.2270\pm0.0008$ & 0.227 & +0.01 & 0.228 & +0.20 \\
$\theta_{23}^q \,(10^{-2})$ & $4.194\pm0.041$ & 4.197 & +0.06 & 4.205 & +0.26 \\
$\theta_{13}^q \,(10^{-3})$ & $3.70\pm0.08$ & 3.701 & +0.02 & 3.711 & +0.13 \\
$\delta_{CKM}$ & $1.139\pm0.023$ & 1.139 & +0.00 & 1.140 & +0.03 \\
\midrule

\multicolumn{6}{c}{\textbf{Neutrino sector}} \\
\midrule
$\Delta m^2_{21} \,(10^{-5}\,\mathrm{eV}^2)$ & $7.49\pm0.19$ & 7.490 & -0.00 & 7.454 & -0.19 \\
$\Delta m^2_{31} \,(10^{-3}\,\mathrm{eV}^2)$ & $2.534^{+0.025}_{-0.023}$ & 2.536 & +0.08 & 2.536 & +0.06 \\
$\sin^2\theta_{12}$ & $0.307^{+0.012}_{-0.011}$ & 0.306 & -0.10 & 0.306 & -0.08 \\
$\sin^2\theta_{23}$ & $0.561^{+0.012}_{-0.015}$ & 0.558 & -0.80 & 0.560 & -0.77 \\
$\sin^2\theta_{13}$ & $0.02195^{+0.00054}_{-0.00058}$ & 0.0219 & -0.04 & 0.0219 & -0.05 \\
\midrule

\multicolumn{6}{c}{\textbf{Baryon asymmetry}} \\
\midrule
$\eta_B\,(10^{-10})$& $6.12\pm 0.04$ &7.1& - & 7.9 & -  \\
\midrule

$\chi^2$ & -- & -- & 0.88 & -- & 0.97 \\
\bottomrule
\end{tabular}
}
\caption{Experimental values (Exp.) of the observables at the $M_Z$ scale, together with their $1\sigma$ uncertainties, are taken from Refs.~\cite{Antusch:2025fpm} for the charged-fermion sector and Refs.~\cite{NUFIT,Esteban:2020cvm} for the neutrino sector. Although the central value of $\sin^2\theta_{23}$ and its $1\sigma$ uncertainty are reported in the table, our numerical analysis accounts for the full allowed range, including both $\theta_{23} < 45^\circ$ and $\theta_{23} > 45^\circ$ solutions; see Ref.~\cite{NUFIT}. The fitted values corresponding to \textbf{BM I} (\textbf{BM II}) are presented in the third (fifth) column, with the associated pulls given in the fourth (sixth) column. The pull of baryon asymmetry $\eta_B$ is not quoted since it is not taken into account in the $\chi^2$ function.
}
\label{tab:fit}
\end{table}

Moreover, we perform a detailed numerical analysis of baryogenesis via leptogenesis~\cite{Fukugita:1986hr}, employing the same method as in our previous work~\cite{Babu:2024ahk,Babu:2025wop}. 
In the fitting procedure, baryon-to-photon number density today $\eta_B$ is first computed using an analytical approximation to make sure the value is close to the observed one $\eta_B^{\rm obs} = (6.12 \pm 0.04) \times 10^{-10}$~\cite{Planck:2018vyg}. Then, a more precise value is numerically determined using flavor-covariant formalism~\cite{Fong:2021xmi} assuming zero initial $N_i$ abundance. For this reason, $\eta_B$ is not taken into account in the $\chi^2$ function.

In the following, we consider two scenarios: \textbf{BM I}, corresponding to the tree-level case of Eq.~\eqref{eq:mnu}, and \textbf{BM II}, which includes one-loop corrections to the neutrino mass matrix as in Eq.~\eqref{eq:corrected_mnu}. The fit parameters of Eqs.~\eqref{eq:MU}--\eqref{eq:MR} at $M_{\rm GUT}$ are presented in Appendix \ref{app:fit_parameters}.

The fitted values of the observables, together with their corresponding pulls, are presented in Table~\ref{tab:fit}, which shows excellent agreement with experimental data. For these benchmark points, the predictions for various observables are listed in Table~\ref{tab:prediction}, including light neutrino masses, neutrinoless double beta decay parameters, the leptonic CP phase, heavy neutrino mass scales, and proton decay branching ratios~\cite{FileviezPerez:2004hn,Nath:2006ut,Saad:2026ucf}.

\begin{table}[th!]
\centering
\footnotesize
\resizebox{0.56\textwidth}{!}{
\begin{tabular}{lcc}
\toprule

\textbf{Observable} & \textbf{BM I} & \textbf{BM II} \\
\midrule

\multicolumn{3}{c}{\textbf{Neutrino masses}} \\
\midrule
$m_1\,(\mathrm{meV})$ & 0.0415 & 0.0459 \\
$m_2\,(\mathrm{meV})$ & 8.654 & 8.634 \\
$m_3\,(\mathrm{meV})$ & 50.360 & 50.354 \\
\midrule

$m_{\beta\beta}\,(\mathrm{meV})$ & 3.708 & 3.689 \\
\midrule

\multicolumn{3}{c}{\textbf{Leptonic observables}} \\
\midrule
$\delta_{\mathrm{PMNS}}\,(\mathrm{deg})$ & 66.777 & 64.048 \\
\midrule

\multicolumn{3}{c}{\textbf{Heavy neutrino masses}} \\
\midrule
$M_1/10^{4}\,(\mathrm{GeV})$ & 5.280 & 5.057 \\
$M_2/10^{12}\,(\mathrm{GeV})$ & 3.547 & 2.257 \\
$M_3/10^{14}\,(\mathrm{GeV})$ & 10.175 & 6.390 \\

\midrule
\multicolumn{3}{c}{\textbf{Proton decay branching ratios (\%)}} \\
\midrule
$\mathrm{BR}(p\to \pi^0 e^+)$ & 44.805 & 44.713 \\
$\mathrm{BR}(p\to \pi^0 \mu^+)$ & 0.367 & 0.375 \\
$\mathrm{BR}(p\to K^0 e^+)$ & 0.586 & 0.580 \\
$\mathrm{BR}(p\to K^0 \mu^+)$ & 3.289 & 3.241 \\
$\mathrm{BR}(p\to \eta^0 e^+)$ & 0.043 & 0.042 \\
$\mathrm{BR}(p\to \eta^0 \mu^+)$ & $3.487 \times 10^{-4}$ & $3.560 \times 10^{-4}$ \\
$\mathrm{BR}(p\to \pi^+ \bar{\nu})$ & 49.894 & 50.030 \\
$\mathrm{BR}(p\to K^+ \bar{\nu})$ & 1.016 & 1.018 \\

\bottomrule
\end{tabular}
}
\caption{Predicted values of selected observables for the benchmark points, including light neutrino masses, the effective neutrinoless double beta decay parameter, leptonic CP phase, heavy neutrino masses, and proton decay branching ratios.}
\label{tab:prediction}
\end{table}

\begin{figure}[th!]
\centering
\includegraphics[width=0.6\textwidth]{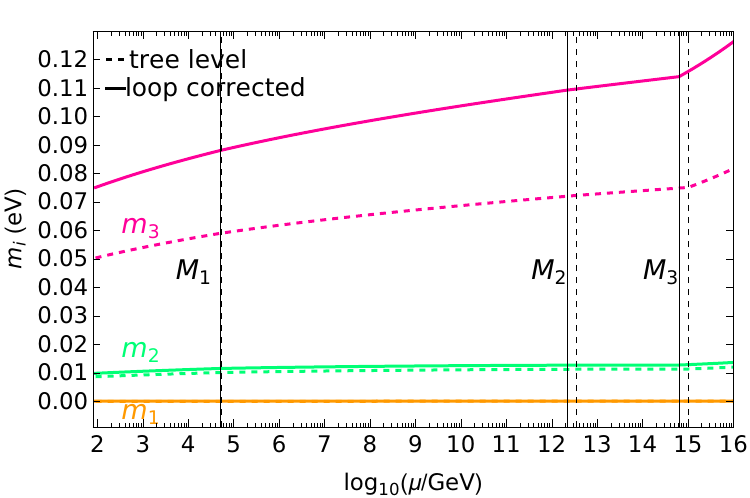}
\includegraphics[width=0.6\textwidth]{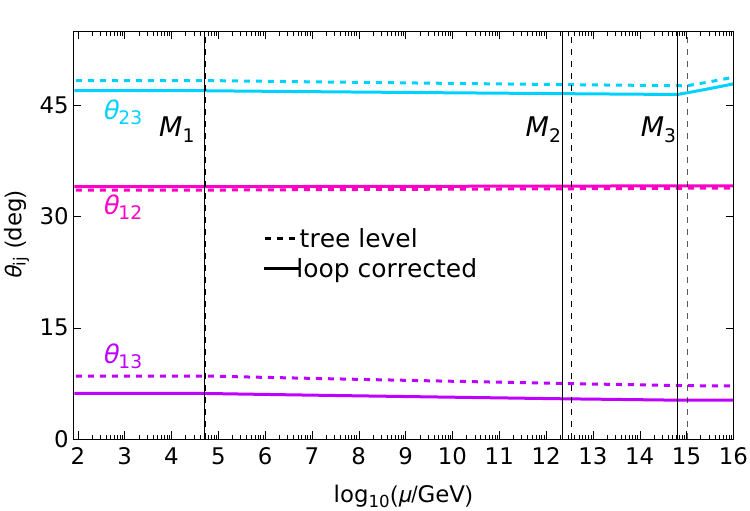}
\caption{  The evolution of neutrino masses and mixing angles under RGE from the GUT scale down to the $M_Z$ scale where the dashed (solid) lines correspond to \textbf{BM I} (\textbf{BM II}). 
} \label{fig:RGE}
\end{figure}

\begin{table}[th!]
\centering
\footnotesize

\textbf{\large Importance of One-Loop Corrections} \\[2mm]
\resizebox{0.8\textwidth}{!}{
\begin{tabular}{lccc}
\toprule
\textbf{Observable} & \textbf{tree-level} & \textbf{One-loop} & \textbf{Relative change (\%)} \\
\midrule

$m_1$ (meV) & $4.15\times10^{-5}$ & $2.83\times10^{-5}$ & \color{blue}$-31.8$ \\
$m_2$ (meV) & $8.65\times10^{-3}$ & $7.22\times10^{-3}$ & \color{blue}$-16.5$ \\
$m_3$ (meV) & $5.04\times10^{-2}$ & $3.49\times10^{-2}$ & \color{blue}$-30.6$ \\
\midrule

$\theta_{12}$ (deg) & $33.57$ & $32.68$ & \color{blue}$-2.64$ \\
$\theta_{23}$ (deg) & $48.34$ & $50.32$ & \color{blue}$+4.08$ \\
$\theta_{13}$ (deg) & $8.51$ & $11.75$ & \color{blue}$+38.06$ \\
\midrule

\bottomrule
\end{tabular}
}
\caption{Case study of \textbf{BM I}. Comparison of neutrino observables between  tree-level fit and after adding one-loop corrects to this existing fit solution. The relative change induced by radiative corrections is defined as $(\text{new} - \text{old})/\text{old} \times 100\%$.  }
\label{tab:loop_corrections}
\end{table}

To illustrate the importance of radiative corrections, we present the \textbf{BM I} predictions for neutrino observables in Table~\ref{tab:loop_corrections}. Recall that \textbf{BM I} corresponds to a tree-level fit. When one-loop corrections are included on top of this tree-level result, the modified values of the observables, together with their relative percentage shifts, are also shown in Table~\ref{tab:loop_corrections}. The resulting corrections are sizeable, clearly demonstrating the impact of radiative effects on neutrino observables. 
Although derived in a concrete setup, our findings are not model-specific and are applicable to a wide class of GUTs with a type-I seesaw structure.

Finally, Figure~\ref{fig:RGE} depicts the evolution of neutrino masses and mixing angles under RGE running. In this figure, the dashed (solid) lines correspond to \textbf{BM I} (\textbf{BM II}). The running is performed from the GUT scale down to the $M_Z$ scale. It is important to recall that, for \textbf{BM II}, the one-loop corrections are incorporated only after running down to the $M_Z$ scale. This figure again highlights the importance of incorporating loop corrections in reliably obtaining the correct model parameters.

\section{Conclusions}\label{sec04}
In this work, we have investigated fermion mass fits in $SO(10)$ GUTs beyond the tree-level approximation by systematically including one-loop finite corrections to the neutrino mass matrix. We find that parameter regions which provide excellent fits to all fermion masses and mixing observables at tree level can nevertheless undergo substantial shifts once radiative corrections are taken into account. This is explicitly demonstrated in a representative benchmark scenario (see Table~\ref{tab:loop_corrections}), where one-loop effects induce sizeable corrections in neutrino observables, in particular leading to sizeable corrections of $\mathcal{O}(30\%)$ in neutrino masses, 
and reaching up to nearly $\mathcal{O}(40\%)$ in $\theta_{13}$.

These results show that loop-induced effects can significantly alter the leptonic sector predictions and therefore critically impact the predictive power of fermion mass fits. Within the $SO(10)$ framework, where all fermion masses and mixings originate from a common set of Yukawa couplings, such effects are inherently global: corrections in the neutrino sector induce correlated shifts across quark and charged lepton observables, thereby reshaping the viable parameter space. Our findings highlight a limitation of conventional tree-level analyses and underscore the strong sensitivity of neutrino observables to higher-order corrections in unified frameworks. In view of the increasing precision of current neutrino oscillation data and the expected improvements from future experiments, a consistent and reliable exploration of the $SO(10)$ parameter space requires the inclusion of one-loop corrections. 
More generally, this inclusion is crucial for the assessment of the viability and predictivity of unified models of fermion masses that feature type-I seesaw mechanism.

\subsection*{Acknowledgments}
CSF acknowledges the support by Fundacão de Amparo à Pesquisa do Estado de São Paulo (FAPESP) Contract No. 2019/11197-6  and Conselho Nacional de Desenvolvimento Científico e Tecnológico (CNPq) under Contract No. 304917/2023-0.  SS also acknowledges the financial support
from the Slovenian Research Agency (research core funding No.\ P1-0035 and N1-0321). The authors acknowledge the Center for Theoretical Underground Physics and Related Areas (CETUP* 2024) and the Institute for Underground Science at Sanford Underground Research Facility (SURF) for hospitality and for providing a conducive environment where this work was initiated.  Additionally, SS acknowledges the Federal University of ABC (UFABC) for its hospitality during the completion of this work. Finally, the authors are grateful to Diego Aristizabal Sierra for discussion.

\appendix

\section{Benchmark fit parameters}\label{app:fit_parameters}

Here we present the benchmark fit parameters for \textbf{BM I} (tree-level neutrino mass) and \textbf{BM II} (tree-level plus one-loop finite corrections for neutrino mass) at $M_{\rm GUT}$.
For \textbf{BM I}, the parameters at $M_{\rm GUT}$ are:
\begingroup
\allowdisplaybreaks
\begin{align}
	&
	\left(\hat p, q, r, s\right) =
	\left(
	3.57629\times 10^{-3}\,e^{i\,0.96833},\;
	1.87989\,e^{i\,1.71829},\;
	2.24823\,e^{-i\,1.57054},\;
	1.2588\times 10^{13}
	\right),  \label{eq:fit01a} 
	\\
	&\frac{\mathcal{M}_2^\mathrm{diag}}{\rm{GeV}}= \left(
	\begin{array}{ccc}
		4.19433\times 10^{-9} & 0 & 0 \\
		0 & 0.28181 & 0 \\
		0 & 0 & 85.62220 \\
	\end{array}
	\right),
	\\
	&
	\frac{\mathcal{M}_1}{\rm{GeV}}=
	\begin{pmatrix}
		4.71247\times 10^{-4}\,e^{i\,1.80795} &
		4.58242\times 10^{-3}\,e^{i\,2.70419} &
		5.25098\times 10^{-3}\,e^{-i\,3.11292} \\
		4.58242\times 10^{-3}\,e^{i\,2.70419} &
		5.97754\times 10^{-2}\,e^{-i\,1.72689} &
		5.62916\times 10^{-1}\,e^{i\,2.62677} \\
		5.25098\times 10^{-3}\,e^{-i\,3.11292} &
		5.62916\times 10^{-1}\,e^{i\,2.62677} &
		2.63917\times 10^{-1}\,e^{-i\,2.00164}
	\end{pmatrix},
	\\
	&
	\frac{\mathcal{M}_3}{\rm{GeV}}=\begin{pmatrix}
		0 &
		2.34061\times 10^{-4}\,e^{-i\,2.86131} &
		3.62802\times 10^{-4}\,e^{-i\,2.36231} \\
		2.34061\times 10^{-4}\,e^{i\,0.28029} &
		0 &
		3.10648\times 10^{-1}\,e^{i\,0.91477} \\
		3.62802\times 10^{-4}\,e^{i\,0.77929} &
		3.10648\times 10^{-1}\,e^{-i\,2.22682} &
		0
	\end{pmatrix}. \label{eq:fit01b}
\end{align}
\endgroup
\noindent
For \textbf{BM II}, the parameters at $M_{\rm GUT}$ are:
\begingroup
\allowdisplaybreaks
\begin{align}
	&
	\left(\hat p, q, r, s\right) =
	\left(
	3.65311\times 10^{-3}\,e^{i\,1.01499},\;
	1.68414\,e^{i\,1.71452},\;
	2.01096\,e^{-i\,1.57058},\;
	7.94376\times 10^{12}
	\right),  \label{eq:fit02a}
	\\
	&\frac{\mathcal{M}_2^\mathrm{diag}}{\rm{GeV}}= \left(
	\begin{array}{ccc}
		6.36620\times 10^{-9} & 0 & 0 \\
		0 & 0.28416 & 0 \\
		0 & 0 & 86.21600 \\
	\end{array}
	\right),
	\\
	&
	\frac{\mathcal{M}_1}{\rm{GeV}}=
	\begin{pmatrix}
		4.75662\times 10^{-4}\,e^{i\,1.82645} &
		4.65480\times 10^{-3}\,e^{i\,2.70308} &
		5.38181\times 10^{-3}\,e^{-i\,3.11512} \\
		4.65480\times 10^{-3}\,e^{i\,2.70308} &
		6.03405\times 10^{-2}\,e^{-i\,1.76515} &
		5.64779\times 10^{-1}\,e^{i\,2.61391} \\
		5.38181\times 10^{-3}\,e^{-i\,3.11512} &
		5.64779\times 10^{-1}\,e^{i\,2.61391} &
		2.47047\times 10^{-1}\,e^{-i\,1.92449}
	\end{pmatrix},
	\\
	&
	\frac{\mathcal{M}_3}{\rm{GeV}}=  \begin{pmatrix}
		0 &
		2.34346\times 10^{-4}\,e^{-i\,2.84713} &
		3.65668\times 10^{-4}\,e^{-i\,2.36274} \\
		2.34346\times 10^{-4}\,e^{i\,0.29447} &
		0 &
		3.47395\times 10^{-1}\,e^{i\,0.90367} \\
		3.65668\times 10^{-4}\,e^{i\,0.77885} &
		3.47395\times 10^{-1}\,e^{-i\,2.23792} &
		0
	\end{pmatrix}.\label{eq:fit02b}
\end{align}
\endgroup

\section{Running of lepton parameters}\label{app:lepton_couplings}

Here we show how much Dirac neutrino and charged lepton Yukawa couplings evolve from $M_{\rm GUT}$ to $M_1$ due to RGE running. 
For completeness, we also show the mild running of right-handed neutrino masses in $M_{\nu_R}$. 

For evaluating the one-loop finite corrections in Eq.~\eqref{eq:1-loop}, one has to use the frozen $M_{\nu_D} = Y_{\nu_D} v$ with $v=174$ GeV and $M_{\nu_R}$ at $M_1$ scale, obtained after integrating out all $N_i$ following Ref.~\cite{Antusch:2005gp}. To obtain an accurate determination of baryon asymmetry from leptogenesis, one should take into account RGE evolution. 
Instead of the full running, this effect is taken into account approximately by evaluating quantities associated to $N_i$ using the parameters at $M_i$ scale.

In the following, we will quote $Y_{\nu_D}$, $Y_E = M_E/v$ and $M_{\nu_R}$ at $\mu = \{M_{\rm GUT}, M_3, M_2, M_1\}$.
For \textbf{BM I}, we have at 

\noindent $\mu = M_{\rm GUT}$:
\begin{align}
	&M_{\nu_R} = {\rm diag}
	\left(
5.2801 \times 10^{4},\;
3.5476 \times 10^{12},\;
1.0779 \times 10^{15}
	\right)\; \mathrm{GeV}, \\
	&
	Y_{\nu_D} =
\begin{pmatrix}
2.6830\times 10^{-6}\,e^{i\,1.8079} &
4.0833\times 10^{-6}\,e^{i\,2.7297} &
4.3116\times 10^{-6}\,e^{i\,0.2537} \\
4.8097\times 10^{-5}\,e^{i\,2.7018} &
4.8777\times 10^{-3}\,e^{-i\,3.0726} &
3.1713\times 10^{-2}\,e^{-i\,3.0150} \\
6.4001\times 10^{-5}\,e^{-i\,3.0979} &
2.6853\times 10^{-2}\,e^{i\,0.2700} &
1.4630\,e^{-i\,3.1407}
\end{pmatrix},
	\\&
	Y_E =
\begin{pmatrix}
2.6829\times 10^{-6}\,e^{i\,1.8078} &
2.8149\times 10^{-5}\,e^{i\,2.6237} &
3.3237\times 10^{-5}\,e^{i\,3.0680} \\
2.4223\times 10^{-5}\,e^{i\,2.7975} &
3.5592\times 10^{-4}\,e^{-i\,1.7478} &
9.2068\times 10^{-4}\,e^{-i\,1.1668} \\
2.6942\times 10^{-5}\,e^{-i\,2.9866} &
7.1636\times 10^{-3}\,e^{i\,2.5487} &
6.7150\times 10^{-3}\,e^{-i\,2.1344}
\end{pmatrix}.
\end{align}

\noindent $\mu = M_3$:
\begin{align}
	&M_{\nu_R} = {\rm diag}
	\left(
5.2801 \times 10^{4},\;
3.5475 \times 10^{12},\;
1.0175 \times 10^{15}
	\right)\; \mathrm{GeV}, \\
	&
	Y_{\nu_D} =
\begin{pmatrix}
2.6088\times 10^{-6}\,e^{i\,1.8080} &
3.9698\times 10^{-6}\,e^{i\,2.7295} &
4.1317\times 10^{-6}\,e^{i\,0.2538} \\
4.6750\times 10^{-5}\,e^{i\,2.7018} &
4.7503\times 10^{-3}\,e^{-i\,3.0721} &
3.0401\times 10^{-2}\,e^{-i\,3.0150} \\
5.9541\times 10^{-5}\,e^{-i\,3.0973} &
2.5013\times 10^{-2}\,e^{i\,0.2699} &
1.3623\,e^{-i\,3.1407}
\end{pmatrix},
	\\&
	Y_E =
\begin{pmatrix}
2.6312\times 10^{-6}\,e^{i\,1.8079} &
2.7588\times 10^{-5}\,e^{i\,2.6233} &
3.4020\times 10^{-5}\,e^{i\,3.0684} \\
2.3755\times 10^{-5}\,e^{i\,2.7975} &
3.4858\times 10^{-4}\,e^{-i\,1.7493} &
9.4270\times 10^{-4}\,e^{-i\,1.1664} \\
2.6430\times 10^{-5}\,e^{-i\,2.9862} &
7.0273\times 10^{-3}\,e^{i\,2.5480} &
6.8783\times 10^{-3}\,e^{-i\,2.1343}
\end{pmatrix}.
\end{align}

\noindent $\mu = M_2$:
\begin{align}
	&M_{\nu_R} = {\rm diag}
	\left(
	5.2801 \times 10^{4},\;
	3.5475 \times 10^{12},\;
	1.0175 \times 10^{15}
	\right)\; \mathrm{GeV}, \\
	&
	Y_{\nu_D} =
	\begin{pmatrix}
		2.6098\times 10^{-6}\,e^{i\,1.8081} &
		3.9712\times 10^{-6}\,e^{i\,2.7297} &
		4.1334\times 10^{-6}\,e^{i\,0.2538} \\
		4.6745\times 10^{-5}\,e^{i\,2.7018} &
		4.7503\times 10^{-3}\,e^{-i\,3.0721} &
		3.0401\times 10^{-2}\,e^{-i\,3.0150} \\
		5.9541\times 10^{-5}\,e^{-i\,3.0973} &
		2.5012\times 10^{-2}\,e^{i\,0.2699} &
		1.3623\,e^{-i\,3.1407}
	\end{pmatrix},
	\\&
	Y_E =
	\begin{pmatrix}
		2.6851\times 10^{-6}\,e^{i\,1.8081} &
		2.8160\times 10^{-5}\,e^{i\,2.6235} &
		3.4718\times 10^{-5}\,e^{i\,3.0684} \\
		2.4242\times 10^{-5}\,e^{i\,2.7976} &
		3.5571\times 10^{-4}\,e^{-i\,1.7493} &
		9.6201\times 10^{-4}\,e^{-i\,1.1664} \\
		2.6971\times 10^{-5}\,e^{-i\,2.9861} &
		7.1707\times 10^{-3}\,e^{i\,2.5481} &
		7.0202\times 10^{-3}\,e^{-i\,2.1342}
	\end{pmatrix}.
\end{align}

\noindent $\mu = M_1$:
\begin{align}
	&M_{\nu_R} = {\rm diag}
	\left(
	5.2801 \times 10^{4},\;
	3.5472 \times 10^{12},\;
	1.0175 \times 10^{15}
	\right)\; \mathrm{GeV}, \\
	&
	Y_{\nu_D} =
	\begin{pmatrix}
		2.6098\times 10^{-6}\,e^{i\,1.8080} &
		3.9714\times 10^{-6}\,e^{i\,2.7295} &
		4.1332\times 10^{-6}\,e^{i\,0.2538} \\
		4.6766\times 10^{-5}\,e^{i\,2.7018} &
		4.7519\times 10^{-3}\,e^{-i\,3.0721} &
		3.0412\times 10^{-2}\,e^{-i\,3.0150} \\
		5.9541\times 10^{-5}\,e^{-i\,3.0973} &
		2.5013\times 10^{-2}\,e^{i\,0.2699} &
		1.3623\,e^{-i\,3.1407}
	\end{pmatrix},
	\\[10pt]
	&
	Y_E =
	\begin{pmatrix}
		2.7304\times 10^{-6}\,e^{i\,1.8080} &
		2.8628\times 10^{-5}\,e^{i\,2.6233} &
		3.5304\times 10^{-5}\,e^{i\,3.0684} \\
		2.4650\times 10^{-5}\,e^{i\,2.7975} &
		3.6172\times 10^{-4}\,e^{-i\,1.7493} &
		9.7826\times 10^{-4}\,e^{-i\,1.1664} \\
		2.7426\times 10^{-5}\,e^{-i\,2.9862} &
		7.2916\times 10^{-3}\,e^{i\,2.5480} &
		7.1380\times 10^{-3}\,e^{-i\,2.1343}
	\end{pmatrix}.
\end{align}

\noindent For \textbf{BM II}, we have at 

\noindent $\mu = M_{\rm GUT}$:
\begin{align}
	&M_{\nu_R} = {\rm diag}
	\left(
5.0572 \times 10^{4},\;
2.2573 \times 10^{12},\;
6.8488 \times 10^{14}
	\right)\; \mathrm{GeV}, \\
	&
	Y_{\nu_D} =
\begin{pmatrix}
2.7081\times 10^{-6}\,e^{i\,1.8265} &
4.1403\times 10^{-6}\,e^{i\,2.5966} &
4.4805\times 10^{-6}\,e^{i\,0.3358} \\
4.8887\times 10^{-5}\,e^{i\,2.7121} &
4.9313\times 10^{-3}\,e^{-i\,3.0732} &
3.5673\times 10^{-2}\,e^{-i\,3.0106} \\
6.5562\times 10^{-5}\,e^{-i\,3.0943} &
3.0840\times 10^{-2}\,e^{i\,0.2590} &
1.4730\,e^{-i\,3.1407}
\end{pmatrix},
	\\&
	Y_E =
\begin{pmatrix}
2.7080\times 10^{-6}\,e^{i\,1.8264} &
2.8367\times 10^{-5}\,e^{i\,2.6327} &
3.3641\times 10^{-5}\,e^{i\,3.0771} \\
2.4784\times 10^{-5}\,e^{i\,2.7835} &
3.6017\times 10^{-4}\,e^{-i\,1.7821} &
9.0992\times 10^{-4}\,e^{-i\,1.1794} \\
2.7946\times 10^{-5}\,e^{-i\,3.0056} &
7.1752\times 10^{-3}\,e^{i\,2.5368} &
6.7628\times 10^{-3}\,e^{-i\,2.0841}
\end{pmatrix}.
\end{align}

\noindent $\mu = M_3$:
\begin{align}
	&M_{\nu_R} = {\rm diag}
	\left(
5.0572 \times 10^{4},\;
2.2572 \times 10^{12},\;
6.3902 \times 10^{14}
	\right)\; \mathrm{GeV}, \\
	&
	Y_{\nu_D} =
\begin{pmatrix}
2.6179\times 10^{-6}\,e^{i\,1.8265} &
4.0017\times 10^{-6}\,e^{i\,2.5963} &
4.2563\times 10^{-6}\,e^{i\,0.3361} \\
4.7237\times 10^{-5}\,e^{i\,2.7118} &
4.7785\times 10^{-3}\,e^{-i\,3.0724} &
3.3901\times 10^{-2}\,e^{-i\,3.0107} \\
6.0091\times 10^{-5}\,e^{-i\,3.0935} &
2.8313\times 10^{-2}\,e^{i\,0.2589} &
1.3518\,e^{-i\,3.1407}
\end{pmatrix},
	\\&
	Y_E =
\begin{pmatrix}
2.6448\times 10^{-6}\,e^{i\,1.8265} &
2.7677\times 10^{-5}\,e^{i\,2.6321} &
3.4586\times 10^{-5}\,e^{i\,3.0777} \\
2.4206\times 10^{-5}\,e^{i\,2.7837} &
3.5114\times 10^{-4}\,e^{-i\,1.7847} &
9.3583\times 10^{-4}\,e^{-i\,1.1792} \\
2.7303\times 10^{-5}\,e^{-i\,3.0051} &
7.0106\times 10^{-3}\,e^{i\,2.5359} &
6.9608\times 10^{-3}\,e^{-i\,2.0838}
\end{pmatrix}.
\end{align}

\noindent $\mu = M_2$:
\begin{align}
	&M_{\nu_R} = {\rm diag}
	\left(
	5.0572 \times 10^{4},\;
	2.2572 \times 10^{12},\;
	6.3902 \times 10^{14}
	\right)\; \mathrm{GeV}, \\
	&
	Y_{\nu_D} =
	\begin{pmatrix}
		2.6180\times 10^{-6}\,e^{i\,1.8266} &
		4.0022\times 10^{-6}\,e^{i\,2.5963} &
		4.2566\times 10^{-6}\,e^{i\,0.3362} \\
		4.7249\times 10^{-5}\,e^{i\,2.7119} &
		4.7785\times 10^{-3}\,e^{-i\,3.0724} &
		3.3901\times 10^{-2}\,e^{-i\,3.0107} \\
		6.0091\times 10^{-5}\,e^{-i\,3.0935} &
		2.8313\times 10^{-2}\,e^{i\,0.2590} &
		1.3518\,e^{-i\,3.1407}
	\end{pmatrix},
	\\[10pt]
	&
	Y_E =
	\begin{pmatrix}
		2.6975\times 10^{-6}\,e^{i\,1.8265} &
		2.8229\times 10^{-5}\,e^{i\,2.6319} &
		3.5276\times 10^{-5}\,e^{i\,3.0777} \\
		2.4688\times 10^{-5}\,e^{i\,2.7837} &
		3.5813\times 10^{-4}\,e^{-i\,1.7847} &
		9.5452\times 10^{-4}\,e^{-i\,1.1791} \\
		2.7847\times 10^{-5}\,e^{-i\,3.0051} &
		7.1502\times 10^{-3}\,e^{i\,2.5359} &
		7.0994\times 10^{-3}\,e^{-i\,2.0838}
	\end{pmatrix}.
\end{align}

\noindent $\mu = M_1$:
\begin{align}
	&M_{\nu_R} = {\rm diag}
	\left(
	5.0572 \times 10^{4},\;
	2.2570 \times 10^{12},\;
	6.3902 \times 10^{14}
	\right)\; \mathrm{GeV}, \\
	&
	Y_{\nu_D} =
	\begin{pmatrix}
		2.6180\times 10^{-6}\,e^{i\,1.8266} &
		4.0022\times 10^{-6}\,e^{i\,2.5963} &
		4.2565\times 10^{-6}\,e^{i\,0.3362} \\
		4.7248\times 10^{-5}\,e^{i\,2.7119} &
		4.7784\times 10^{-3}\,e^{-i\,3.0724} &
		3.3900\times 10^{-2}\,e^{-i\,3.0107} \\
		6.0091\times 10^{-5}\,e^{-i\,3.0935} &
		2.8313\times 10^{-2}\,e^{i\,0.2590} &
		1.3518\,e^{-i\,3.1407}
	\end{pmatrix},
	\\[10pt]
	&
	Y_E =
	\begin{pmatrix}
		2.7392\times 10^{-6}\,e^{i\,1.8272} &
		2.8657\times 10^{-5}\,e^{i\,2.6275} &
		3.5809\times 10^{-5}\,e^{i\,3.0776} \\
		2.4950\times 10^{-5}\,e^{i\,2.7837} &
		3.6335\times 10^{-4}\,e^{-i\,1.7867} &
		9.6886\times 10^{-4}\,e^{-i\,1.1820} \\
		2.8290\times 10^{-5}\,e^{-i\,3.0053} &
		7.2495\times 10^{-3}\,e^{i\,2.5358} &
		7.1920\times 10^{-3}\,e^{-i\,2.0837}
	\end{pmatrix}.
\end{align}

\bibliographystyle{style}
\bibliography{reference}
\end{document}